# Epitaxial GaN Nanorods via Catalytic Capillary Condensation

H. W. Seo and Q. Y. Chen

Department of Physics and Texas Center for Superconductivity & Advanced Materials, University of Houston, Texas, USA

L. W. Tu and C. L. Hsiao

Department of Physics and Center for Nanoscience & Nanotechnology, National Sun Yat-Sen University, Taiwan, Republic of China

M. N. Iliev and W. K. Chu

Department of Physics and Texas Center for Superconductivity & Advanced Materials, University of Houston, Texas, USA

Abstract

Intrinsic catalytic process by capillary condensation of Ga-atoms into nanotrenches, formed among impinging islands during the wurzite-GaN thin film deposition, is shown to be an effective path to growing GaN nanorods without metal catalysts. The nano-capillary brings within it a huge imbalance in equilibrium partial pressure of Ga relative to the growth ambient. GaN nanorods thus always grow out of a holding nanotrench and conform to the boundaries of surrounding islands. The nanorods are epitaxially orientated with $<0001>_{GaN}$ // $<111>_{Si}$ and $<2\underline{11}0>_{GaN}$ // $<110>_{Si}$ similar to the matrix. Concaved geometry is essential and is a condition that limits the axial dimension of the nanorods protruding above the base (matrix) material region. Revelation of the growth mechanism in the current context suggests that fabrication of nano quantum structures with controlled patterns is enabling for any attainable dimensions.

GaN has undergone extensive research and development for the past decade because of its perceived commercial value for blue laser and light emitting diode applications [1, 2]. In these efforts, one of the main challenges was to obtain morphologically smooth thin films, since samples frequently ended in rough surface or columnar structures, which are undesirable defects from a two-dimensional thin film device application perspective. More recently, however, such spurious structures have attracted a great deal of interest, as many believe they may be exploited for controlled formation of nanostructures suitable for quantum device purposes [3]. Indeed, plenty of alluring nano-geometric shapes have been reported, ranging from flexible and often free-standing or entangled nanowires [4--6] to epitaxial spikes of nanorods or nano-columns standing on a base material [7--10]. Such attempts in nanostructure formation, nonetheless, must move from the presently naturally occurring processes to scalable designs such as patterned arrays in order to realize their full application potentials. For this purpose, it would be desirable to understand the mechanism of nanostructure formation from which some process design rules may be inferred.

Previous studies on the growth mechanisms of nanowires of GaN [5] and other semiconductors [7, 11] all more or less rested on the classical catalyst-assisted vapour-liquid-solid (VLS) or vapour-solid (VS) models, established earlier in an effort to understand the growth of whiskers[12]. The nanorods observed in this work, nevertheless, are quite different in that they are always accompanied by nanotrenches [13--15] and require no extrinsic metallic catalysts as the nuclei. These phenomena have not been recognized [9, 10] prior to this work and we argue that capillary condensation of Ga-atoms within the nanotrenches serves as a catalytic agent to trigger the nanorod growth.

The objective of this work is to elucidate this point of view by modelling the growth mechanism based on the observed nanorods in relation to their nanotrench precursors. Our interest in this study was mainly on the wurtzite structure (0001) GaN epitaxial films populated with nanorods. Underneath such samples typically is a template buffer layer deposited at a lower temperature (500-600 °C) under high nitrogen to gallium ratio conditions. The GaN films were grown by plasma assisted MBE on <111> oriented silicon substrates, as reported earlier [16]. The nanorods, with lateral dimensions ~10-100 nm and longitudinal dimensions ~0.4-0.6 μm above the film surface, always coexist with a "nanoflower" pattern [16], and are largely rooted at the vicinity of the homo-epitaxial buffer layer to Si interface alongside the nanotrench, as shown in Fig.1. The nanoflower is essentially a V-shaped crater of six-fold symmetry, with its bottom connected to a trench which extends through the entire film down to the silicon interface. Similar geometry on the close packed planes of some other materials of either cubic symmetry [17], or wurtzite structure, such as SiC [18], have also been observed. The nanorods are pin-shaped, by and large, with tip pointed toward the bottom, and are detached from the walls of the nanotrenches typically by a lateral gap of several nanometres. These nanorods, nanotrenches, and their accompanying nanoflowers are randomly dispersed, sometimes clustering around different regions of the sample.

The origin of nanotrench formation has hitherto mostly been attributed in the literature to the threading dislocation, as suggested by Frank [19]. In this context, a dislocation of large Burgers vector may lower its energy by emptying the highly strained core region. In balance, the critical core radius is $r = Gb^2 / 8\pi^2\gamma$ [19], where $G$ is the shear modulus, $b$ is the Burgers vector, and $\gamma$ is the surface energy of the material. (Here the planes of relevance are the $\{10\underline{1}0\}$ family). Unfortunately, the observed core radii (2*r* ~ 10-100 nm) are too large per core energy [13--15]. Suppose that ***b*** is along <0001> and $b$ = 0.518 nm (the lattice constant along the c-axis), $G = C_{13} \sim 10^{11}$ N/m$^2$ [2], and $\gamma \sim 1.89$ J/m$^2$ [20], then we have 2*r* ~ 0.34 nm, suggesting that the existence of such open cores is not driven by the strain-field of a dislocation, otherwise this would have given $\gamma \sim$ 0.006 J/m$^2$, which is out of range by far for a solid compound.

In order to account for the above mentioned discrepancy, we suggest that there are two types of nanotrenches. Type-I trench, having smaller diameter (~1 nm) and containing no nanorods, can indeed be the trace of an open-core dislocation [21] situated at centre of an island. Type-II trench, as referred to in this letter simply as the nanotrench, on the other hand, is typically one hundred times larger. Earlier reports showed that this type of trench could either be empty [13, 15], or filled with inversion domains [14, 15]. Within this work, we realize that the type-II nanotrenches come to exist as a result of coarsening hexagonal islands. One may ask why it would favour island growth that leads to the nanotrenches in lieu of a smooth film since, in classical views [22], lattice mismatch strain are relieved beyond a critical thickness through the generation of misfit dislocations. Recently, it was recognized that strain relief can also occur instead by way of surface roughening [23], as demonstrated in single layer thin films or superlattices of Group-IV compounds [24] and III-V compounds [25]. Surface roughening relaxes mismatch strain at the cost of higher total surface energy, giving a thermodynamically lower energy state as compared to that of a smooth, strained, and possibly concaved surface. This roughened surface is a precursor to the island or columnar structure. In the end the strain field would be unevenly distributed and the trough regions of the roughened surface should be more strained than their crest counterparts. This is in agreement with our earlier observations at the GaN buffer to substrate interface [16]. For GaN on AlN, it is known that the lattice mismatch $\delta = (a_{GaN} - a_{AlN}) / a_{AlN}$ is ~ + 2.4% and the critical thickness, beyond which surface roughening takes place, is only 1-3 mono layers [25]. Hence island formation for GaN on Si in our samples should start even earlier at the interface region since there $\delta = (a_{GaN} - a_{Si}) / a_{si}$ ~ -17%. This unusually large lattice mismatch might have been reconciled to a larger coincidence lattice with respect to the Si substrate, though this hypothesis remains to be verified.

The nucleation and growth process, we reckon, start with three randomly chosen precursor nuclei on the vertices of an irregular triangle (Fig.2 (a)). Consequently, a voided-region of equilateral triangle would develop as illustrated in Fig. 2 (b-1) when the islands encounter each other. Certainly, on some occasions, the islands may leave no voided regions behind if the relative positions of the nuclei happen to be along $[1\underline{1}0]_{Si}$ or $[2\underline{11}0]_{GaN}$ directions. But in any case, in the simplest scenario, the voided triangular areas are thermodynamically unstable, unlike islands of metallic elements for which triangular form is common [26]. For growth that occurs on the close-packed {111} plane of a Si lattice, we hence suggest that a transformation would take place to reduce the surface energy by corner filling via some self-regulated surface diffusion along the edges where the islands wet the substrate. This eventually results in a hexagonal nanotrench, as sketched in Fig. 2 (c-1). Such concept was vaguely touched upon by J. Elsner *et al* [27].

With the hexagonal empty region in place, the islands would continue to grow in the vertical dimension and the six facets surrounding the region would then be elevated, eventually becoming what one observes as a six-leaf nanoflower. The nanotrenches underneath the nanoflower are essentially like an attached capillary tube (Fig. 2(d)). When the capillary tube is long enough, plausibly at the end of low temperature buffer layer growth when the aspect ratio of the tube is high, capillary condensation of Ga atoms occurs as a consequence of the decreases in the equilibrium vapour pressure due to the reduced radius of curvature of a concave surface, as seen from the Thomson (Lord Kelvin) equation $k_BT\log[P_{Ga}(r)/P_{Ga}(\infty)] \sim -2\gamma^{LV}V_{Ga}/r$ [28]. For a convex surface, conversely, the negative sign is reversed and a small radius of curvature would favour evaporation. As a cross examination, for 100 nm, 1 μm, and 10 μm tubes, the ratio $P(r) / P(\infty)$ are, respectively, $\sim 5x10^{-24}$, $5x10^{-3}$ and 0.6 with molar volume for Ga being $\sim 11.8\ cm^3/mol$. The flat surface equilibrium partial pressure $P_{Ga}(\infty) \sim 10^{-8}$ torr at 500-600 °C for the buffer layer and $\sim 10^{-5}$ torr at 720 °C for the ultimate film. To estimate the critical radius r* for a Ga condensate, below which no sustainable growth may be commenced, one may assume a spherical geometry of the Ga nano-droplet with point contact to the silicon substrate, then from $r^* = 2\gamma^{LV}/3\Delta G_v$, taking into account the reduction of energy due to the above mentioned capillary effect, where $\Delta G_v$ is the volume free energy upon condensation taken as the heat of fusion ~5.59 kJ/mol., one finds $2r^* \approx 6$ nm; this may well account for the observed smallest sizes of nanorods, which are about 10 nm in diameter.

GaN nanorod grows faster along <0001> via the canonical VLS mechanism [12] by reaction of the nitrogen plasma with Ga liquid clusters while its surrounding GaN islands also grow alongside to form the base materials, though at a slower rate. Figs. 2 (c)-(g) illustrate this sequence of evolution. The cross-sectional views indicate that the lateral dimensions of nanorods and nanotrenches are graded in the beginning and there is a gap in-between, as shown in Figs. 2 (d), (e) and (f). Here, we show when capillary condensation starts and nanorod eventually sprouts from the nanotrench due to its higher growth rate. As the nanorod outgrows the nanotrench and starts to stick out of the base film surface, an intermediate case between Fig. 2(e) and (f), the equilibrium vapor pressure would gradually increase which favours the evaporation of the droplets on top of the nanorod (Fig. 2 (f)). The process eventually reaches a steady state after which the rods grow at the same rate as the surrounding film via VS growth mechanism since the conditions for capillary condensation no longer exist; otherwise the nanorods would grow enormously long and thus stand far above the base material, which is not what was observed, as suggested in Fig. 2(g). According to our observation, as indicated by the top-views in Fig. 3 (a)-(c), where the SEM surface morphologies of three samples are shown, at the beginning stages of island formation and impingements, i.e. for the buffer layer, only nanotrenches but no nanorods exist. As higher temperature growth commences, nanorods begin to emerge as seen in Fig. 3 (b). This coincides with the above reasoning that capillary condensation doesn't occur until the nanotrench is deep enough to foster the process. Moreover, the size of a nanorod depends on that of the holding nanotrench, while that of the nanotrench, in turn, depends on the density of the impinging islands with relevant lateral dimensions (also on order of 10-100 nm). As seen from the evolving morphology, both nanorod and matrix islands coarsen with time at a constant temperature as seen by comparing Fig. 3(c) with 3 (b).

The nanorod density also depends on the growth temperature, though in a more complex way than simply being proportional to the island density. For instance, direct growth without low-temperature buffer layer would foster columnar structure and suppress nanorod growth, which is consistent with the concept of nano-capillary condensation as at high temperature there are less regions of coalescence where the radius of curvature is small enough—most are flat boundaries formed by impinged nano- or submicron-columns. Sporadic nanorods are observable at those corner sites where the radius of curvature meets the criterion of capillary condensation [29]. This suggests that while the matrix most likely would follow the traditional thermal mechanism of grain- or island growth (including columnar structures), nanorod growth is highly restricted by the seeding of Ga atoms in the nano-capillary. Furthermore, the shape of nanorod confirms to that of the holding nanotrench and can, in reality, be less symmetric than a perfect hexagon depicted above. As shown in the four SEM micrographs of Fig.4, we see some intermediate states of a geometrical evolution of the nanorods. These geometrical variants may be related to the asymmetric corner filling of the triangular basins, as alluded to earlier, at the beginning of nanotrench formation, thus also further providing a convincing evidence for the conformity of nanorods to their holding nanotrenches.

The growth mechanism described above hinges on the roughened surface under which capillary tubes are developed to serve as the nucleation sites for the nanorods. Since surface roughing originates from the lattice mismatch strain, then can variations in lattice mismatch yield a different surface morphology and hence alter the formation of nanorods or nanotrenches? While an exact correlation is yet to be established, it is compelling, in our view, that nanorods indeed tend to form under the circumstance of a larger lattice mismatch. This is consistent with our observation from the top-view SEM micrographs (Fig. 3) that the nanorods would grow larger as the matrix grows thicker until it reaches an optimal size and then remains constant. Here, as the film grows thicker or if it grows at higher temperatures, the extent to which strain-relief occurs also expands until the island growth and strain relaxation run their courses. At that moment, the mound-like islands would finally take shape. The slant of these islands, and hence the ultimate sizes of the nanotrenches they enclose, is a manifestation of the lattice mismatch between GaN and its Si substrate at the film growing temperature. Beyond that point the relaxation process is complete and the sizes of nanorods would stop growing.

Note other factors such as Ga:N ratio and the method of film growth can also play an intricate role on the nanostructure formation. Lower temperature deposition under N-rich condition favours higher nucleation density [21] while slower growth rates give more time for condensation and nitrogen atom diffusion in the droplet. In this vein, it may be understandable why no nanorod formation has been observed so far for other methods of inherently higher growth rate operating at much higher temperature such as metallorganic chemical vapour deposition (MOCVD). Certainly, this warrants a more careful investigation, even though N-rich plasma-assisted MBE growth at medium temperatures of 750-800 °C truly appears to be most favourable for the GaN nanorods to form [8--10, 16] as long as the growth starts with a low temperature deposited buffer layer. The nanorods are epitaxially orientated, based on electron backscatter Kikuchi patterns, with rod axis // <0001> of the base area and <111>$_{Si}$ while the closed packed <2$\bar{1}\bar{1}$0> axis for both the nanorods, nanotrenches and the base film regions are all parallel to the <110> close packed direction of the {111} oriented silicon substrate [30]. Such high degree of epitaxy implies strong enough molecular interacting forces dictating the atomic arrangements of the nanorods with respect to their surrounding matrix. The matrix is Ga-polar, judged from its resistance to the $H_3PO_4$ etching as compared to the nanorods [31], which all disappeared after etching. However, due to the flimsiness of nanorod as bound by the nanotrench, we can't conclude the nanorods to be N-polar [30].

In summary, we report that nanotrenches catalyze the formations of GaN nanorods through the capillary condensation of Ga-atoms. This implies that self-assembled extrinsic metal catalysis, frequently used to produce nanowires and nanorods, can be replaced with nano capillaries. The capillaries reported here are the nanotrenches which arise from impinging islands during the film growth. The concept of nano-capillary condensation is inspiring as it can be extended to integrated fabrications of designed patterns for useful electron or photonic devices with the help of contemporary advanced lithographic means.

This work was supported in part by the Welch Foundation and in part by the State of Texas through the Texas Centre for Superconductivity and Advanced Materials at the University of Houston. Work at National Sun Yat-Sen University was supported by the National Science Council of the Republic of China through the Center for Nanoscience & Nanotechnology.

**Figure Captions:**

**Fig 1: SEM picture of GaN Nanorods:**

**(a) GaN nanorods protruding out of nanoflower craters (20 º tilt view).**

**(b) Cross-sectional view.**

**(c) Top view.**

**Fig 2: Evolution of nanotrench and nanorod:**

**a) Initial stage of GaN island growth.**

**b) Impinging hexagonal islands and the formation of a triangular void region.**

**c) Corner filling of the triangular void and its evolution into hexagonal shape, precursor to the nanoflower.**

**d) Evolution of the nanoflower and start of the nanotrench formation and capillary condensation of Ga atoms in the trench.**

**e) VLS growth mechanism prevails and the nanorod grows faster, leading to protrusion above the nanoflower. As the protrusion occurs, the condition for capillary condensation diminishes and VS growth mechanism takes over.**

**f) The ultimate structure.**

**Fig .3) Surface morphologies showing various stages of nanorod formation: (a) Buffer layer 37 nm grown at 590 º C, (b) 50 nm film at 720 ºC on buffer layer (a), and (c) 150 nm film at 720 ºC on buffer layer (a).**

**Fig.4) SEM top view showing different geometric shapes assumed by the nanorods, manifesting the intermediate stages of corner filling mentioned in the text.**

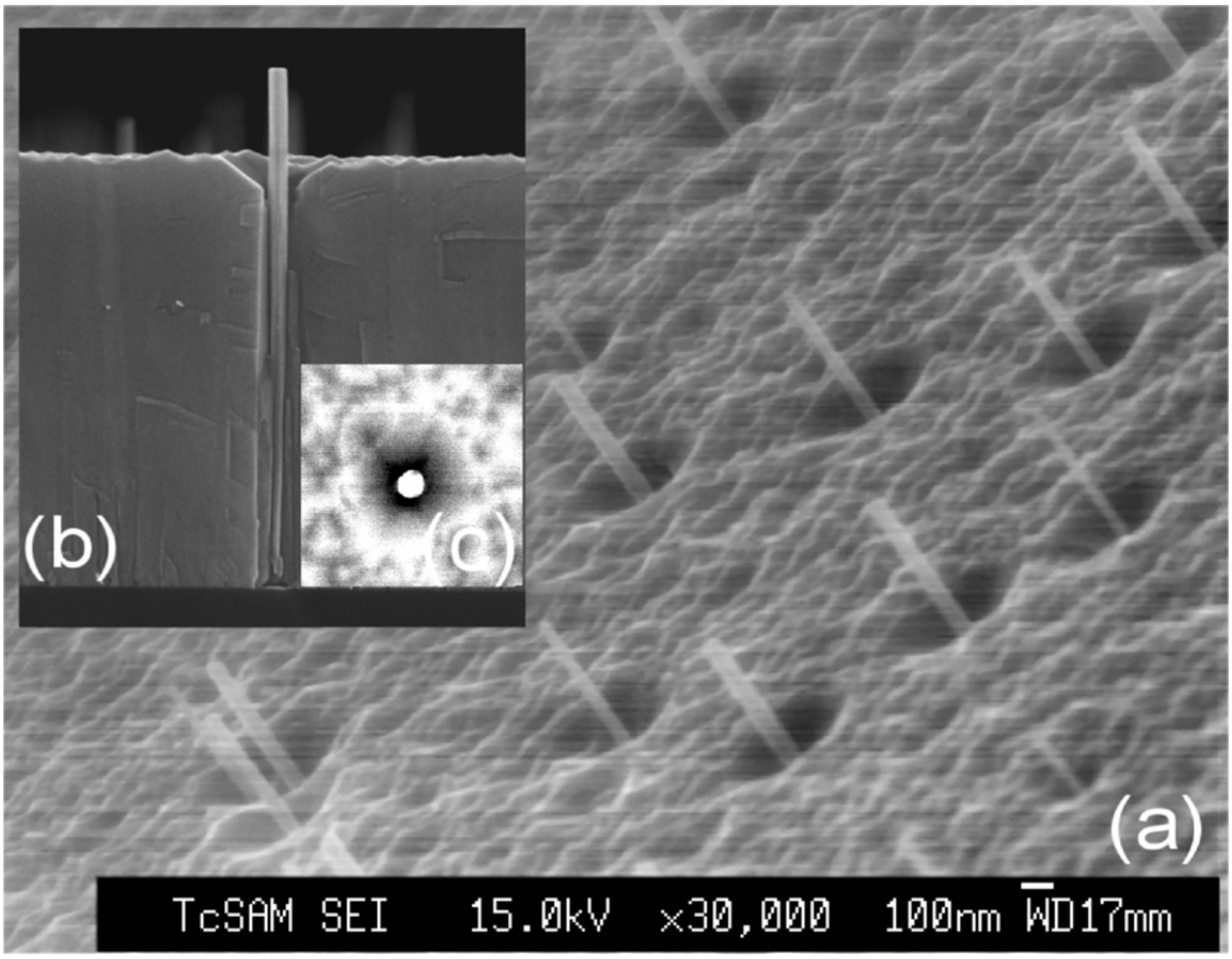

[ Fig. 1]

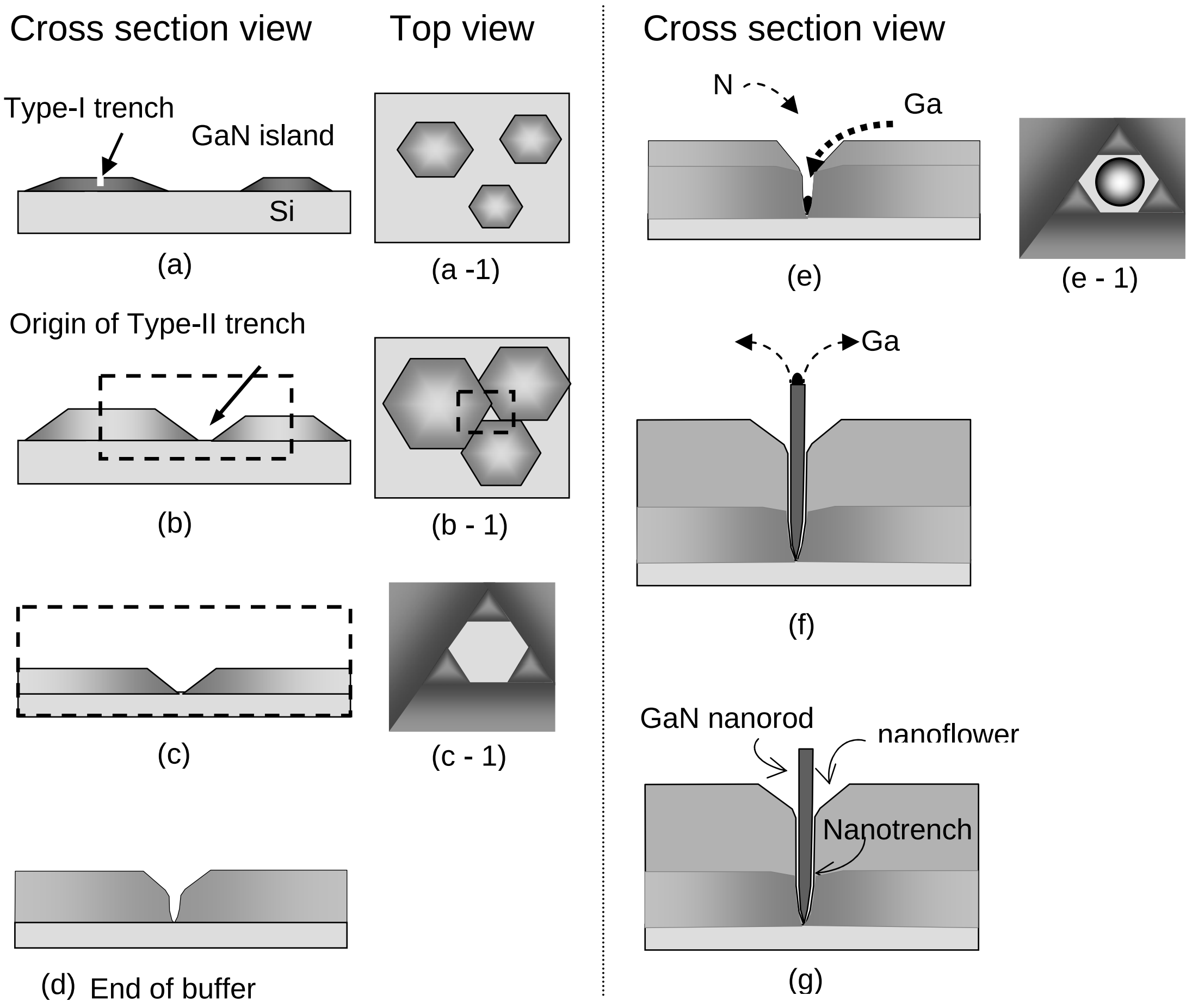

[ Fig. 2]

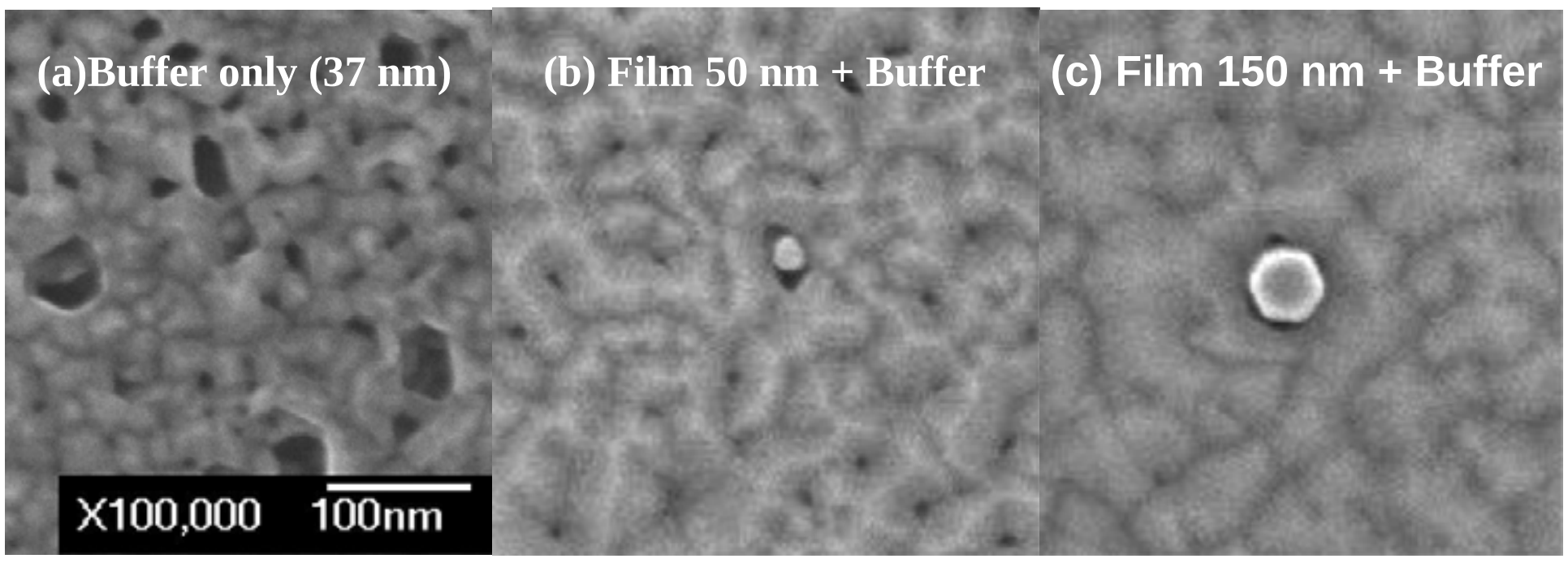
(a)Buffer only (37 nm)
(b) Film 50 nm + Buffer
(c) Film 150 nm + Buffer
X100,000
100nm

[Fig. 3]

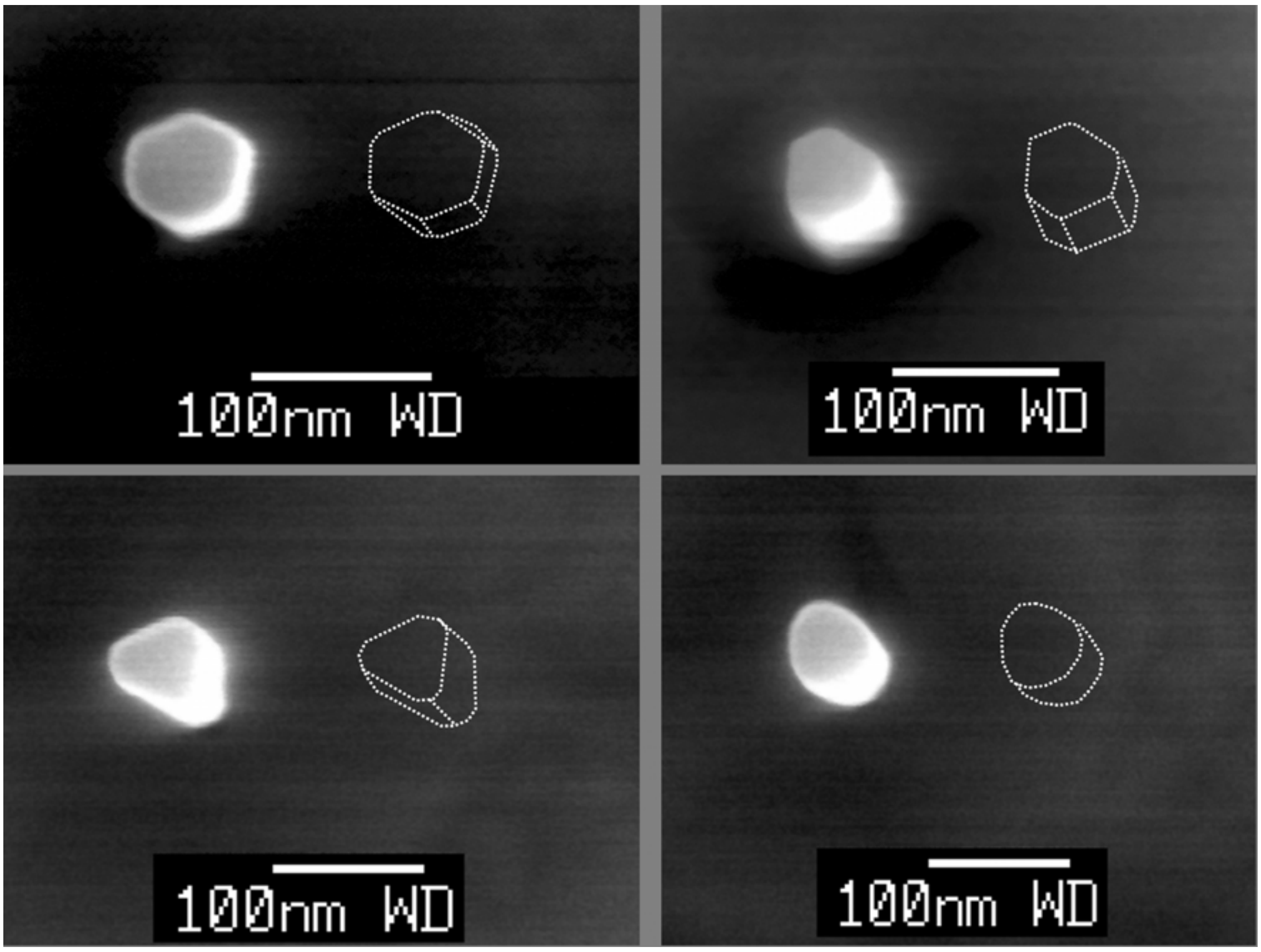

[ Fig. 4]